\documentclass[pdflatex,sn-vancouver-ay]{sn-jnl}
\usepackage{graphicx}
\usepackage{multirow}
\usepackage{xcolor}
\usepackage{textcomp}
\usepackage{amsmath}
\usepackage{longtable}
\usepackage{geometry}
\usepackage{array}

\title{\textbf{Diachronic and synchronic variation in the performance of adaptive machine learning systems: The ethical challenges}}

\author[1]{\fnm{Joshua} \sur{Hatherley}}\email{jjh@hum.ku.dk}

\author[2]{\fnm{Robert} \sur{Sparrow}}

\affil[1]{\orgdiv{Center for the Philosophy of AI}, \orgname{University of Copenhagen, Denmark}}

\affil[2]{\orgdiv{School of Philosophical, Historical, and International Studies}, \orgname{Monash University, Australia}}

\date{April 1st, 2025}

\abstract{\textit{Objectives:} Machine learning (ML) has the potential to facilitate ‘continual learning’ in medicine, in which an ML system continues to evolve in response to exposure to new data over time, even after being deployed in a clinical setting. In this paper, we provide a tutorial on the range of ethical issues raised by the use of such ‘adaptive’ ML systems in medicine that have, thus far, been neglected in the literature.

\textit{Target audience:} The target audiences for this tutorial are the developers of machine learning AI systems, healthcare regulators, the broader medical informatics community, and practicing clinicians. 

\textit{Scope:} Discussions of adaptive ML systems to date have overlooked the distinction between two sorts of variance that such systems may exhibit -- diachronic evolution (change over time) and synchronic variation (difference between cotemporaneous instantiations of the algorithm at different sites) -- and under-estimated the significance of the latter. We highlight the challenges that diachronic evolution and synchronic variation present for the quality of patient care, informed consent, and equity, and discuss the complex ethical trade-offs involved in the design of such systems. 

\bigskip

This is a pre-print of: Hatherley, Joshua and Robert Sparrow. 2023. Diachronic and synchronic variation in the performance of adaptive machine learning systems: The ethical challenges. \textit{Journal of the American Medical Informatics Association} 30(2): 361-366. \href{https://doi.org/10.1093/jamia/ocac218}{10.1093/jamia/ocac218}}

\begin{document}

\maketitle

\section{Introduction}

Machine learning (ML) has the potential to facilitate ‘continual learning’ in medicine, in which an ML system continues to adapt and evolve in response to exposure to new data over time, even after being deployed in a clinical setting. Leveraging this ‘adaptive’ potential of medical ML could generate significant benefits for patient health and well-being. Recent engagements with the ethical issues generated by the use of adaptive ML systems in medicine have typically been limited to discussions of ‘the update problem’: how should systems that continue to change and evolve post-regulatory approval be regulated? In this paper, we draw attention to an important set of ethical issues raised by the use of adaptive machine learning systems in medicine that have, thus far, been neglected and are highly deserving of further attention. 

Discussions of adaptive machine learning systems to date have overlooked the distinction between two sorts of variance that such systems may exhibit — diachronic evolution (change over time) and synchronic variation (difference between cotemporaneous instantiations of the algorithmic system at different sites) — and under-estimated the significance of the latter. Both diachronic evolution and synchronic variation will complicate the hermeneutic task of clinicians in interpreting the outputs of AI systems, and will therefore pose significant challenges to the process of securing informed consent to treatment. Equity issues may occur where synchronic variation is permitted, as the quality of care may vary significantly across patients or between hospitals. However, the decision as to whether to allow or eliminate synchronic variation involves complex trade-offs between accuracy and generalisability, as well as a number of other values, including justice and non-maleficence. In some contexts, preventing synchronic variation from emerging may only be possible at the expense of the wellbeing, and the quality of care available to, particular patients or classes of patients. Designers and regulators of adaptive ML systems will need to confront these issues if the potential benefits of adaptive ML in medical care are to be realised.

\section{Adaptive machine learning in medicine}\label{2}

ML is a form of AI that involves “programming computers to optimize a performance criterion using example data or past experience” \citep[3]{alpaydin2020introduction}. The application of ML in medicine could significantly improve the delivery of medical care, and expand the availability of medical knowledge and expertise, among other benefits \citep{esteva2019guide,rajkomar2019machine,rajpurkar2022ai,sparrow2019promise}. ML systems can be either ‘locked’ or ‘adaptive’. Locked ML systems have parameters fixed prior to clinical deployment, and do not continue to learn from new data over time. While, to date, regulatory approvals of medical AI systems have been limited to locked systems the U.S. Food and Drug Administration (FDA) is considering regulatory approval for adaptive ML systems, which evolve as they are exposed to new data (“continuous learning”), even after the system has been deployed in a clinical setting \citep{usfood2018artificial,usfood2021artificial}. We will refer to these sorts of ML devices as (Medical) Adaptive Machine Learning System(s) (MAMLS).

The use of MAMLS could have a number of benefits for patients. In some applications, MAMLS can continuously ‘tune’ their algorithms to individual patients’ physiology, along with any changes that occur in a patient’s physiology over their lifetime, thereby contributing to the realisation of ‘personalised medicine’. The use of ML to deliver personalised medicine is already being explored via the combination of ML with a variety of other new and emerging technologies \citep{banaei2019machine}. For example, ML-enabled wearables and implantables have been developed to enable personalised identification of ventricular arrythmias and hypoglycaemic events for diabetic patients, and also to predict the onset of seizures in patients with drug-resistant epilepsy \citep{porumb2020precision,jia2020personalized,cook2013prediction,pinto2021personalized}.

Additionally, MAMLS could be trained on data collected from particular cohorts of patients to tune their performance to the features of the cohorts of each particular clinical site or institution \citep{ong2023prediction}. For example, MAMLS could be used to predict risk of hospital readmission for outpatients, or to identify patients at a high-risk of heart attack within particular communities \citep{yu2015predicting}. Some researchers are already seeking to enable such site-specific training of medical ML systems by making the source codes of their algorithms freely available online \citep{hong2018predicting}.

\section{The update problem}\label{3}

While there has been some engagement with the ethical issues raised by MAMLS, it has mostly been confined to discussions of ‘the update problem’. Existing regulatory approaches in healthcare and medicine were designed to address products that do not evolve over time, such as pharmaceuticals. Consequently, the capacity for ongoing evolution in MAMLS presents a serious challenge for regulators. As Babic and co-authors have written: “After evaluating a medical AI/ML technology and deeming it safe and effective, should the regulator limit its authorization to market only the version of the algorithm that was submitted, or permit marketing of an algorithm that can learn and adapt to new conditions?” \citep[1202]{babic2019algorithms}. If they approve MAMLS, regulators may be exposing patients to risks that have developed in the system post-deployment. However, restricting regulatory approvals to locked systems places a strong limit on the potential benefits that ML could generate for patient health outcomes. 

In their recent Proposed Regulatory Framework for Modifications to ML-Based Software as a Medical Device (SaMD) \citep{usfood2018artificial} and subsequent Artificial Intelligence/Machine Learning (AI/ML)-Based Software as a Medical Device Action Plan \citep{usfood2021artificial}, the US FDA has attempted to address the update problem. A key feature of the FDA’s preferred approach is the requirement for manufacturers of MAMLS to provide Algorithmic Change Protocols (ACPs) as part of their applications for pre-market approval. ACPs are supposed to outline how a MAMLS will change over time and what the limits of these changes will be. They will also require manufacturers to state how they will mitigate any risks that these changes will present. The FDA suggests that this approach could allow MAMLS to be approved and deployed in clinical settings without a need for ongoing regulatory review. 

A number of serious criticisms have been raised against the FDA’s proposed framework \citep{babic2019algorithms,gerke2020need}. For instance, the proposal gives little indication as to how the performance of MAMLS will be monitored in practice, even suggesting that manufacturers could monitor these systems themselves. We are sympathetic to many of the concerns that have been expressed in the literature. However, we believe that the current focus on regulatory challenges that MAMLS present has led researchers to overlook the broader set of ethical issues that the use of these systems in medicine will present.

\section{Two types of variation: Diachronic and synchronic}\label{4}

The literature on the ethics of MAMLS is cognizant that these systems will evolve over time – a phenomenon that we shall call diachronic evolution. As MAMLS continue learning from new data, their parametric weightings will change from update to update. They will respond differently to identical input data at different times. Their accuracy and performance will evolve over time, for better or worse. They may even adopt different classes of algorithmic bias as they continue to learn and evolve. 

However, it is less often recognised that variation will emerge between copies of a MAMLS that have been implemented across different sites. 

Synchronic variation refers to the differences that will emerge between copies of a MAMLS implemented at different sites or in different patients. MAMLS will be deployed across diverse clinical settings with different data collection policies, organisational procedures, user behaviours, data infrastructures, and patient demographics, each of which will affect the datasets upon which these systems learn. Even small variations in the datasets on which an algorithm learns can have significant effects on what it learns. If each copy of a MAMLS learns from data collected from the site at which it has been deployed, either exclusively, or even just to fine tune its parameters after initial learning from a training dataset, then these differences between site-specific datasets mean that copies of a MAMLS deployed at different sites (or devices implanted in different individuals) are likely to diverge over time. Eventually, identical data entered into different copies of a MAMLS will likely cause these systems to generate different outputs. 

\begin{figure}
    \centering
    \includegraphics[scale=0.88]{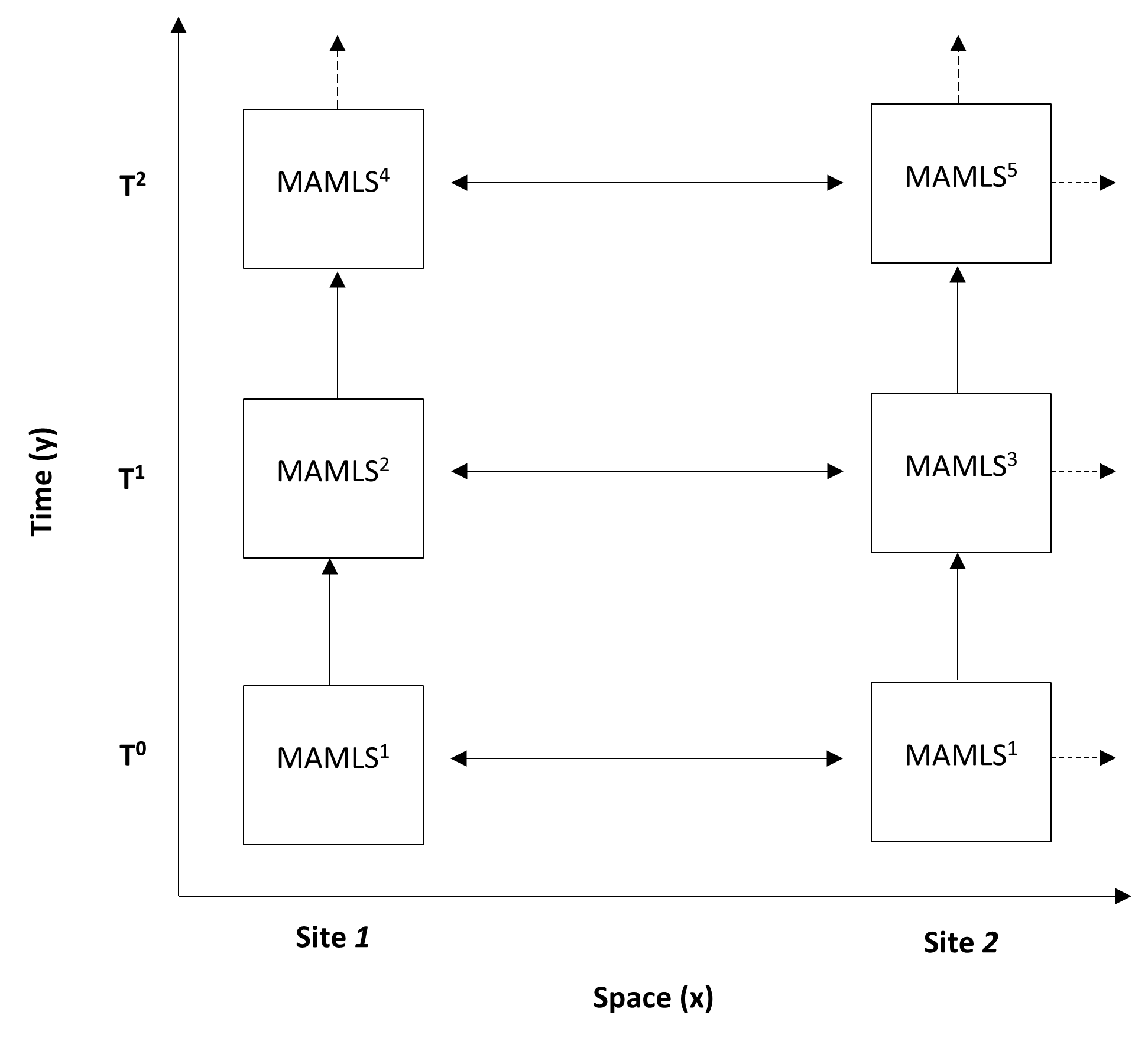}
    \caption{Schematic representation of diachronic evolution (y-axis) and synchronic variation (x-axis) in a MAMLS deployed across two sites.}
    \label{fig:fig1}
\end{figure}

Figure \ref{fig:fig1} illustrates the relation between these two types of variation.

Diachronic evolution alone would seem to require that regulators monitor the evolution of each MAMLS over time and conduct post-market surveillance of the product’s performance, errors, and instances of patient harm as it evolves. The possibility of synchronic variation suggests that, in addition regulators should monitor each copy of a MAMLS, due to the gradual divergence that may occur between each copy of the product over time. The additional administrative burden that this could entail would be expensive and time-consuming for both manufacturers, purchasers, and regulatory agencies. Moreover, the more instantiations there are, the more likely it is that one will go catastrophically wrong and undercut support for all of them. These factors could increase costs for manufacturers and chill the incentive to develop these systems in the first place. 

\section{Federated learning to prevent synchronic variation}\label{5}

In some cases, however, manufacturers may have the option of eliminating synchronic variation entirely via the adoption of ‘federated learning’ (FL), which “involves training statistical models over remote devices or siloed data centers, such as mobile phones or hospitals, while keeping data localized” \citep[50]{li2020federated}. To date, interest in FL in healthcare has mostly been driven by their potential to maintain privacy \citep{rajpurkar2022ai,usynin2021adversarial}. However, if FL can be used to train MAMLS, they could allow each iteration of a MAMLS to learn on the same pool of data, which would preclude the emergence of synchronic variation. 

Eliminating synchronic variation would have a number of benefits for stakeholders. For instance, it would reduce the burden of regulators and manufacturers by eliminating divergence and variation between copies of a MAML and thus the risk that different copies might require distinct regulatory evaluation and approval. In some cases, it might improve the generalisability of MAMLS by enabling these systems to learn from larger, more heterogenous datasets collected across multiple clinical sites. As we will argue below, it could also eliminate the potential for inequities in standards of care to emerge across clinical sites. 

However, the decision to allow or eliminate synchronic variation carries practical trade-offs (along with some ethical trade-offs: see section \ref{6.6}). Implementing FL in MAMLS may require updates to existing digital infrastructure that may be prohibitively expensive for many clinics and hospitals. Federated learning, for instance, “require[s] investment in on-premise computing infrastructure or private-cloud service provision and adherence to standardised and synoptic data formats so that ML models can be trained and evaluated seamlessly” \citep[4]{rieke2020future}. Moreover, if data collected at one hospital cannot easily be transferred to new sites without first being processed, this processing may itself introduce a form of synchronic variation by virtue of adding different extra layers of code to the AI system at different sites. Where implanted medical devices include MAMLS, FL will only be possible if these devices can transmit the result of training on local data back to other instantiations of the learning algorithm and can update the local algorithm in the light of the results of the training of other versions there-of, which increases the risk to patient privacy and of iatrogenic harm, including as a result of hacking. 

In an important set of cases, then, users will face a choice between either allowing synchronic variation to occur, or not using MAMLS at all.

\section{Ethical considerations}

The deployment and use of MAMLS generates a number of ethical concerns relating to the quality of patient care and doctor-patient relations, informed consent to treatment, threats to health equity and problems of obsolescence, and harms to particular cohorts of patients. 

\subsection{Impact on quality of care}

The use of MAMLS presents a number of risks to the quality of patient care. 

Left to continue learning post-deployment MAMLS may adopt erroneous and potentially dangerous associations from new input data that could jeopardise patient health. In one well-known case, a mortality-prediction ML system learned to classify asthmatic patients presenting to the emergency department with pneumonia as ‘low risk’ due to a true but misleading correlation in the training data \citep{caruana2015intelligible}. If operationalised, the system would have presented critical risks to patient safety. With MAMLS there is a risk that such errors will emerge post-deployment as a result the process of continual learning. Moreover, MAMLS are susceptible to the phenomenon of ‘catastrophic forgetting’, in which a MAMLS overwrites what it has previously learned during the process of learning from new data, leading to sudden poor performance that could significantly jeopardise the quality of physician judgments and the health and safety of patients \citep{van2019three}. Finally, MAMLS are susceptible to hacking and adversarial attacks, including by ‘data poisoning’. In locked ML systems, adversarial attacks can only affect individual outputs \citep{finlayson2019adversarial}. In MAMLS, however, adversarial attacks could interfere with the performance of the system (or systems) in all future uses. These possibilities highlight the urgency of the ‘update problem’.

\subsection{Challenges to clinical interpretation}

Furthermore, achieving downstream benefits from the use of ML in medicine is critically dependent upon clinicians’ ability to understand, interpret, and act on the outputs of these systems \citep{hatherley2024virtues}. For instance, clinicians must decide how much epistemic weight they ought to give the outputs of algorithms in their clinical decision-making. Placing too much or too little weight on the outputs of an algorithmic system can result in patient harm, even death. An example of clinicians placing too little epistemic weight in the output(s) of an algorithmic system is ‘alert fatigue’, which refers to “declining clinician responsiveness to a particular type of alert as the clinician is repeatedly exposed to that alert over a period of time, gradually becoming ‘fatigued’ or desensitized to it” \citep[e145]{embi2012evaluating}. Alert fatigue can lead clinicians to ignore important alerts, potentially resulting in patient harm or death \citep[for a particularly egregious instance of patient harm caused by alert fatigue, see][]{wachter2017digital}. An example of patient harm caused by clinicians placing too much epistemic weight in the outputs of an algorithmic system is ‘automation bias’, which “refers to errors resulting from the use of automated cues as a heuristic replacement for vigilant information seeking and processing” \citep[47]{mosier2017automation}. The presence of diachronic evolution and synchronic variation in MAMLS will pose a significant challenge to clinicians being able to reliably interpret and act upon the outputs of these systems. If every time clinicians encounter a MAMLS it is a subtly (or occasionally not so subtly) different system – different both to previous iterations, and between patients and across clinical sites – it may be exceedingly difficult for them to be confident how it is functioning and how much they should trust it. These challenges are further complicated by the opacity of ML systems, which make it difficult to understand how or why a system works or has produced a certain output \citep{hatherley2020limits}. 

Admittedly, that the performance of MAMLS will change over time, and will differ between sites and/or patients, does not in-and-of-itself distinguish them radically from other systems with which clinicians must engage in the course of their professional practice. Clinicians who work across different institutions often have to take account of differences in the way things are done, or particular devices are set up, in each institution. The fact that diagnostic tools and treatments are evolving all the time is, after all, why continuing medical education is so important. However, the key virtue of MAMLS is their ability to continuously improve at a faster rate than existing diagnostic tools without human intervention or oversight. The speed with which MAMLS evolve may outpace clinician’s abilities to adapt to these changes.

\subsection{Impact on doctor-patient relations}

Where clinicians make use of MAMLS for the purpose of clinical decision-support, both diachronic evolution and synchronic variation will pose challenges to communication between doctors and patients and reduce the capacity for shared decision making. If clinicians are themselves not able to understand precisely what has changed between each update to a MAMLS system, or how, precisely, the system they are dealing with at this site, or in this patient, differs from other iterations, they may struggle to identify and explain the factors that are casually relevant to their ultimate decision about a diagnosis and/or treatment plan. In particular, they may find it difficult to provide the patient with counter factual information that might be relevant to shared decision-making about treatment. Importantly, this effect may occur even if the clinician is in fact justified – and can explain to the patient that they are justified – in relying on the MAMLS because of its superior accuracy relative to the alternatives.

It is sometimes argued that the use of AI and ML could allow clinicians more time to spend with their patients \citep{israni2019humanizing,topol2019deep}. However, the various tasks associated with maintaining AI and ML systems could equally lead to increased administrative burdens for clinicians that could further interfere with the quality of care and empathy in the doctor-patient relationship \citep{maddox2019questions,sparrow2020high}. This risk seems particularly acute in the case of MAMLS, because healthcare institutions will likely need to significantly expand the scope of their data collection policies and procedures to be able to provide the continuous stream of new data that training MAMLS will require. 

The potential of MAMLS to evolve over time may also be expected to exacerbate the issues related to computers being ‘the third party in the room’ in clinical consultations. As Christopher Pearce and others have noted, the introduction of computers into healthcare settings has transformed what was originally a dyadic relationship, between the doctor and patient into a triadic relationship between the doctor, the patient, and the doctor’s computer \citep{pearce2011patient,pearce2020consulting}.  Both doctor and patient now spend some – perhaps even much – of their time ‘together’ looking at and relating to the computer: information provided by the computer shapes the course of the consultation. If the doctor’s computer is – or accesses – a MAMLS then this will add an important temporal dimension to the relationship between the doctor and the computer and the patient and the computer. What the computer ‘says’ may change over time. This alone may be sufficient to draw more of the doctor’s and the patient’s attention to the computer. However, the fact that the operations and the outputs of the MAMLS may change also opens up the possibility that doctors will become involved in trying to manage or shape those changes in order to meet their, and their patients, expectations. One might imagine clinicians trying to influence the evolution of the MAMLS by curating the data that they input into it, in the same way many of us now try to manage the recommendation engines of Spotify or Netflix. Clinicians’ relationships with MAMLS will evolve along with the MAMLS and we should expect that at least some clinicians may want to be active in shaping the former evolution – and, thus, the latter.

\subsection{Challenges to informed consent}

Insofar as it is not typically considered necessary for clinicians to inform patients about the technologies that they have used to inform their clinical recommendations, the use of MAMLS for decision support is unlikely to have implications for informed consent. However, where MAMLS assist in the delivery of medical treatment (e.g. robotic surgery or, hypothetically, AI guided radiation therapy) their nature may well be relevant to the process of securing informed consent to treatment. 

The use of ML in treatments already involves new risks that may need to be disclosed to patients, such as the threat of cyberattack \citep{finlayson2019adversarial,kiener2021artificial}. Adaptive learning will introduce additional risks, including the risk of catastrophic forgetting and of algorithmic biases developing post-deployment, which may need to be disclosed to patients in order to allow them to make an informed decision of the use of such systems. Moreover, the potential of MAMLS to evolve and to differ between sites and patients means that the provision of general or ‘standard’ information about treatments guided by MAMLS may not be sufficient to secure informed consent to treatment. A patient who returns to a medical clinic for treatment involving a MAMLS after some time will undergo treatment that may differ subtly, or even significantly, from that they received in their previous visit. Similarly, a patient who moves from one hospital to another, which has implemented a version of the same MAMLS, may be subject to different levels of risk – indeed, different risks – in each location. Fully informed consent, then, may require that patients are made aware of the risks associated with treatment by the particular MAMLS that is involved in their treatment. However, diachronic evolution and synchronic variation, coupled with the characteristic opacity of ML systems, mean that it may not always be possible for manufacturers to provide information about the specific risks associated with a particular iteration of a MAMLS. 

\subsection{Equity and obsolescence}

One hopes that, with appropriate regulation, the continuous learning of MAMLS will lead to improved outcomes for patients over time. In-and-of itself, then, diachronic evolution in the performance of MAMLS should not raise issues of equity.

The ability of MAMLS to adapt to specific patient cohorts and improve the performance of the system amongst these cohorts has the potential to promote health equity by better serving the needs of minority groups that are often under-represented in the training data used to train locked models. However, where synchronic variation is permitted, it is also possible that the difference in the performance of MAMLS at different sites or in different patients may become so pronounced as to generate serious issues of justice in relation to the quality of healthcare available to different cohorts. Particular instantiations of a MAMLS may have biases that are more pronounced, more numerous, or more consequential within the patient cohort that they serve, than other iterations of the product deployed at different sites. Moreover, it is possible that some iterations of a MAMLS product may become stuck in local minima during the learning process, such that their performance stagnates while others continue to improve. In some cases, these performance disparities may become so large that the MAMLS available to particular sites/patients are effectively obsolete.

\subsection{Costs to particular cohorts}\label{6.6}

The challenges that synchronic variation present for equity may serve as another incentive for manufacturers and regulators to try to eliminate synchronic variation. However, although FL is likely to enhance the generalisability of MAMLS, as Futoma and co-authors have noted, “the demand for universal rules – generalisability – often results in [ML] systems that sacrifice strong performance at a single site for systems with mediocre or poor performance at many sites” \citep[e489]{futoma2020myth} \citep[see also][]{burns2020machine}. Disease, symptoms, side-effects, and so on occur with differing probabilities across lines of race, sex, gender, ability, and so on, and the application of a one-size-fits-all model across different subpopulations will often result in a system having differing utility for members of different cohort. Indeed, it can result in a model that is sub-optimal for all groups, or optimal only for the dominant subpopulation – a phenomenon known as ‘aggregation bias’ \citep{suresh2019framework}. For this reason, the decision to prevent synchronic variation in MAMLS involves an ethical and political trade-off between prioritising the health and well-being of dominant groups and the prioritisation of the health and well-being of marginalised groups. 

\section{Conclusion}

We have argued that the implementation of MAMLS raises a number of challenging ethical issues that have thus far received little attention. We distinguished between two sorts of variance that such systems may exhibit — diachronic evolution (change over time) and synchronic variation (difference between cotemporaneous instantiations of the algorithm at different sites). Diachronic evolution complicates the hermeneutic task of clinicians and could interfere with downstream patient health benefits. Maintaining the digital infrastructure and data collection requirements necessary to enable continual learning in MAMLS may generate greater administrative burdens for human physicians, resulting in compromised relations of care and empathy between doctors and patients. Synchronic variation has the potential to generate inequities between clinical sites using the same MAMLS. The choice between site-specific and FL approaches involves a trade-off between pursuing generalisability or local impact, and may be to the detriment of particular cohorts of patients. These ethical issues require sustained attention if we are to realise the benefits of continuous learning in medicine.

\section*{Funding statement}

This research was supported by the Australian Government through the Australian Research Council's Centres of Excellence funding scheme (project CE140100012). RS is also an Associate Investigator in the ARC Centre of Excellence for Automated Decision-Making and Society (CE200100005) and contributed to this paper in that role. JH was supported by an Australian Government Research Training Program scholarship. The views expressed herein are those of the authors and are not necessarily those of the Australian Government or Australian Research Council.

\bibliography{main}
\end{document}